# SMART HOME ENVIRONMENT MODELLED WITH A MULTI-AGENT SYSTEM

Mohammad RASRAS[1], Iuliana MARIN[2], Șerban RADU[3]

*A smart home can be considered a place of residence that enables the management of appliances and systems to help with day-to-day life by automated technology. In the current paper is described a prototype that simulates a context-aware environment, developed in a designed smart home. The smart home environment has been simulated using three agents and five locations in a house. The context-aware agents behave based on predefined rules designed for daily activities. Our proposal aims to reduce operational cost of running devices. In the future, monitors of health aspects belonging to home residents will sustain their healthy life daily.*

**Keywords**: smart home, multi-agent system, K-Nearest Neighbor algorithm, K-Means Clustering algorithm

**1. Introduction**

Smart home, also known as an intelligent house, incorporates special devices that manage house features. The technology of a smart home uses appliances that can communicate among each other and are used to build an environment where many features are automated. In addition, a smart home is a residence place that is equipped with sensors and actuators, and contains smart objects, including a network that allows the exchange of information among existing objects and occupants. An Internet connection is used to connect the smart house with the outside world.

In a context-aware environment, the management of appliances and systems is adopted, such as lighting and heating, in order to help with day-to-day life by automated technology. This can ensure security, comfort, entertainment, and convivence. The communication between devices and appliances of this environment is a unique ecosystem that enhances the life quality of the occupants in a smart home, by monitoring and controlling the residents' activities. Research area in this domain is focused on developing ubiquitous solutions, which depend on collecting information from residents and environment [1], as such environment can play a major role in enhancing the quality of the resident's life. The main services that a smart home can offer are convivence, energy utilization, secure environment, and health assistance [2]. By living in a world of devices, recent research also considered the management of municipal solid waste for a better recycling [3], renewable energy from solar panels [4], and in the healthcare domain, namely illnesses assessment, such as Alzheimer [5].

[1] Eng., Dept. of Engineering in Foreign Languages, University POLITEHNICA of Bucharest, Romania, e-mail: se_mohammad_rasras@yahoo.com
[2] Eng., Dept. of Engineering in Foreign Languages, University POLITEHNICA of Bucharest, Romania, e-mail: marin.iulliana25@gmail.com
[3] Eng., Dept. of Computer Science, University POLITEHNICA of Bucharest, Romania, e-mail: serban.radu@upb.ro

Our motivations in the current research cover several factors. Firstly, convenience is the most important factor, because it is related to time. For example, when an inhabitant walks inside the smart home, a process of managing lights and other appliances is running, with minimum response of time. Secondly, health monitoring is an important aspect of our research, because the suggested design ensures that some health parameters are observed, like toileting-frequency and sleeping patterns. Moreover, managing energy consumption is also taken into consideration.

The proposed design of the smart home can reduce the wasted energy on devices, namely lights and TVs. Finally, privacy for building a non-intrusive system is taken care of, by respecting the private life of occupants.

The paper is structured as follows. Section 2 outlines the literature research regarding agents and multi-agent systems. Section 3 describes the environment where the residents' activities are observed using sensors and actuators, while being in five different locations. Energy consumption and health factors are controlled by agents, in order to decrease energy consumption and to determine problems related to sleep quality and toilet usage respectively. In section 4, three test scenarios are explained. The last section presents the conclusions and future work.

## 2. Literature Research

Scientific researchers have given several definitions for agents in the last years. They also agreed that it is hard to give a precise description for the agent term, as the word 'agent' is a general one for an entity in the development and research areas [6]. In paper [7], the term agent is defined as a software component that performs tasks on behalf of their users. On the other hand, the research in paper [8] defines it as a contained program with the capability to control decision-making and to act based on its perception for the surrounding environment.

Software multi-agent systems (MAS) is a collection of heterogeneous working entities that have their problem-solving and interacting abilities among them, in order to reach an agreed goal [9]. Moreover, MAS is a subfield of artificial intelligence that has the goal to provide principles to construct complex systems, including multiple agents and mechanisms used for coordinating the agents' behavior. This field has been rapidly evolving in the distributed artificial intelligence (DAI) field [10]. Systems that include the usage of multi-agents are applied in problems scenarios and solving methods with different perspectives. MAS systems take advantage of interactions patterns, such as cooperation (joint efforts together to achieve a certain goal), coordination (organize problem to solve activities that avoid risky interactions), and negotiation (come to an acceptable agreement) [11]. The necessity for software systems that have multiple agents, which communicate in a peer-to-peer manner, becomes complex and can lead to large-scale problems. To cope with this issue of complexity, abstraction and modularity are needed to develop modular parts (agents), in which a specific aspect of the problem is handled [12]. Applications of MAS are developed considering the autonomy characteristic of agents, in addition to the ability to interact among themselves in a specific high-level protocol language [13].

The work [14] presents a goal-oriented intelligent environment, based on distributed target-oriented computer architecture. This allows users to achieve a high degree of integration in a multi-agent system, minimizing the level of interaction difficulty. To deal with this, they describe how to determine the target properties and the quality-related parameters of the plan in a smart home scenario. The proposal performs in a low scalability threshold, but when it is high, the system shows an increase in its complexity, making the agents slower to perform their tasks. In paper [15], an agent-based smart home system is developed, using TAOM4E and Jade-Leap platform for remote environment monitoring. A central ontology is used to represent context and support behavioral changes, based on emotional user feedback. The input comes from a wireless sensor network that incorporates light, temperature, humidity, and distance sensors.

Research [16] focuses on the concept and implementation of dual digital technology in smart home applications. The work tests basic concepts alongside the challenges to design a smart home. Subsequent sessions are focused on digital twin (DT) concepts and their implementation. The paper compares the results with dual digital and conventional Internet of Things (IoT) implementations. The document describes the problems encountered in smart home automations and how they are fixed with DTs' concepts. The article shows how DT adds a benefit to a smart home. The project is also designed to simulate situations where security is the dominant feature. Therefore, an example of an intruder is successfully tested. Researchers [17] propose a smart home architecture that contains sensors layer to gather information, middle layer to model the environment, and the application layer. Machine learning techniques are applied to the system, in order to make it adaptive. The focus is on the application layer, because of the variety of services that it can offer. In order to cope with the changing behavior of users, an unsupervised temporal differential machine learning algorithm is applied.

Rules for control logic are used to design smart houses. A system based on object-oriented methods and a graph database is designed by a group of researchers to manage the automatic IoT control, along with user behavior [18]. The response time is better than for systems which have the control mode based on ontologies. Such a semantic web approach is tested in paper [19], where energy decentralization is studied, considering an electric vehicle, a heat pump, and home appliances. SPARQL queries are used to gather data related to energy data processing. Home energy optimization is a subject of interest, as it reduces greenhouse gas emissions and supports environmental protection. For this purpose, researchers use a series of microservices, which include artificial neural algorithms [20, 21], smart grids with relays [22], and Cloud containers [23].

All the research studied denote the need to study multiple environment and healthcare factors for people who live in their own houses. The repetitiveness of actions determines patterns which can be observed. Our paper proposes a light-weight solution to manage energy consumption, mostly during the current times when the globe struggles with an energy crisis. In addition, the proposed system is non-intrusive for a large, targeted audience.

### 3. Methodology

#### 3.1. The environment

The proposed smart home is assumed to have two residents, called (X) and (Y), and an analyzer, which is responsible to update, modify, and display information gathered by the system. The environment contains five locations, bedroom1, bedroom2, bathroom1, bathroom2, and a living room. All locations, excepting the living room, are non-shared locations. For instance, resident X is the only one who can enter bedroom1, bathroom1, and the same rule applies for resident Y, regarding bedroom2 and bathroom2. On the other hand, the living room is a shared location, in which both residents can stay and perform activities. In Fig. 1 is a simple representation of the simulated smart home environment.

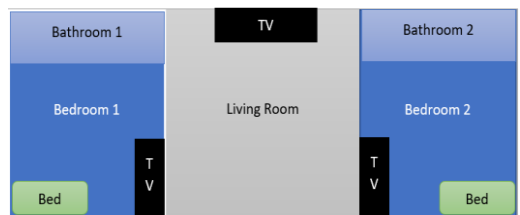

Fig. 1. Simple Representation of the Smart Home

Locations are equipped with sensors that detect motion and brightness, along with actuators, which are also used to control devices. In addition, lights and TVs are influenced by the agents' decisions. Each location is assumed to have one light source. Moreover, three TVs are located in bedroom1, bedroom2, and the living room. All lights are fully powered by the software agents, meaning that they can only be turned on and off, based on agents' decisions. On the other hand, TVs give the residents' preference to turn them on and off. Additionally, they can be turned off by the software agent in two cases, once the residents leave the corresponding location without turning them off, or if the resident falls asleep. The main functional requirements of our system are described in the use-case diagram from Fig. 2.

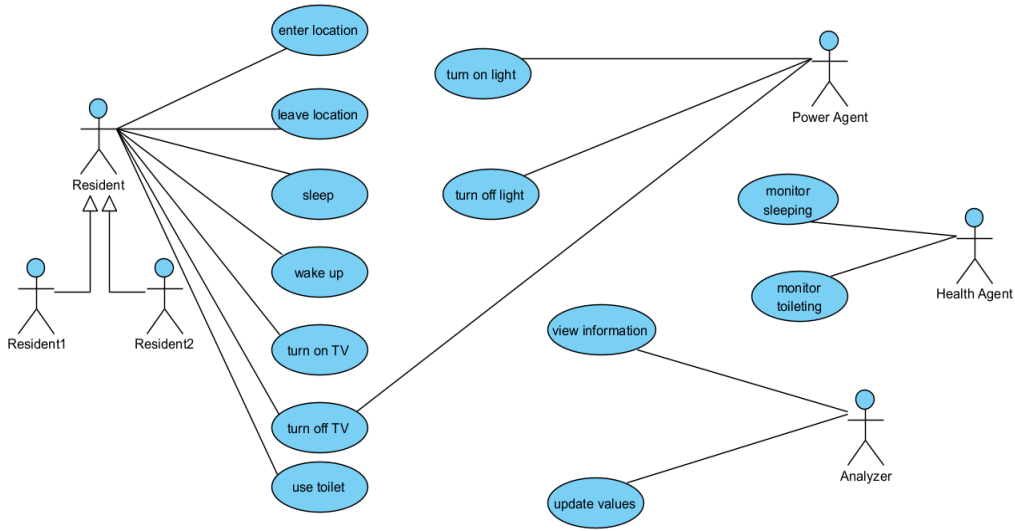

Fig. 2. Use-case Model of the Smart Home

Fig. 2 shows the system actors and the tasks assigned to them. Both the Power and Health Agents are system actors. Residents can enter/leave a location, sleep/wake up in bedrooms, watch TV in bedrooms/living room, and use the toilet in bathrooms. The Analyzer is the actor that represents the health care giver or the administrator of the smart home. It is also responsible to update the sleep quality pattern, delete old data, and view results of the Health Agent's operations.

The smart home environment is simulated using Java desktop application (J2SE). Java is an object-oriented language that allows building a modular application, also supporting concurrent programming. Residents X and Y, along with Analyzer, are implemented as Java threads. Each thread (resident) can browse the permitted locations and can turn on the TV. Residents can do their human needs of sleeping in bedrooms or toileting in bathrooms. On top of Java, Easy Rules engine is used to allow agents to behave based on predefined rules. Easy Rules is simple and powerful to build a software application. It allows to create a collection of objects with conditions, store and iterate them to evaluate conditions and perform actions [24]. The system main features are: to manage energy consumption, by ensuring no waste energy is consumed, to monitor health aspects related to sleeping and toileting factors, and to display details about locations, including energy, occupied time, activities, and their duration.

### 3.2. Managing energy consumption

Managing the energy consumption on lights is carried out by the agent responsible to monitor power (Power Agent). This agent decides to turn on or off the light, based on location occupancy, darkness level, and if residents sleep in bedrooms. Fig. 3 illustrates a sequence diagram that shows the behavior of Power Agent in bedroom. The diagram explains the scenario of an occupant when he/she enters it, after which darkness level is checked, and light is determined to be turned on or off. When the resident turns on the TV, after which he/she sleeps, the agent turns off the devices, and when he/she wakes up, the light is turned on. All the devices are turned off once the resident leaves the current location.

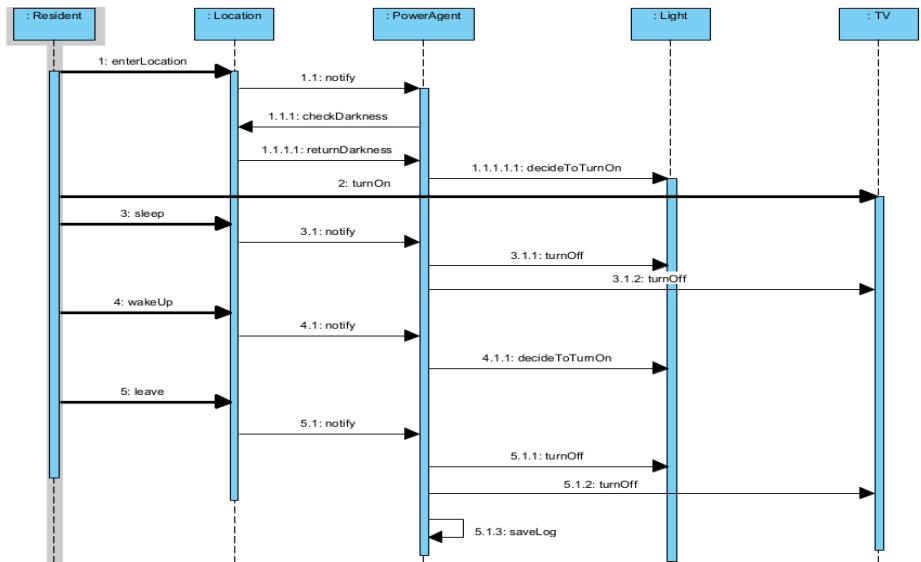

Fig. 3. Behavior of the Power Agent in Bedroom

### 3.3. Monitoring health factors

Our monitored factors for health are sleep quality and toileting frequency. The system stores the resident's sleeping period and saves it. The Health Agent has a predefined pattern of sleeping. This agent compares the period of sleeping with a predefined one and determines and persists the triggered sleeping quality value.

The second factor is achieved by counting the toileting frequency of a resident and persists the value to identify health issues. This is achieved by implementing the K-Nearest Neighbor (KNN) algorithm on a predefined set, to determine if the resident's measures are normal or not. The predefined set of the analyzer is designed with attributes and values stored as labeled data. The attributes are age, weight as person fitness in terms of weight to height ratio, and toileting frequency. The distance needed by the KNN algorithm is used to determine abnormalities.

The distance which is used is the Euclidean one [25], defined as:

$$Distance = \sqrt{(x2-x1)^2 + (y2-y1)^2 + (z2-z1)^2} \qquad (1)$$

Applying Euclidean distance (d) to the predefined labeled data results:

$$d = \sqrt{(P\_Age - G\_Age)^2 + (P\_Weight - G\_Weight)^2 + (P\_Toileting - G\_Toileting)^2} \quad (2)$$

where P is the resident's value, that is predefined and gathered by the system, and G is given for labeled data values.

Lastly, the results are compared, based on the supervised (KNN) and unsupervised (K-Means Clustering, KMC) machine learning techniques. In the case of KMC, the input data set is not split into training and test data, but rather associated between a data point and a cluster, which is not known in advance. Healthcare practitioners or the Analyzer, in our case, can benefit from KMC to identify illness patterns, which are not already known in a data set [26, 27].

## 4. Discussion

Two test categories are carried out to verify the system correctness for the applied scenarios. The first category is related to energy consumption, while the second one is to check results of a resident's health. The tested scenarios take into consideration the virtual environment, for which the simulation includes a specific time interval per activity to certify if the power consumption is kept under control, and the residents' health stays within the normal parameters.

### 4.1. Testing power consumption

The following situation of 10 hours interval is tested, based on residents' behavior, when resident X stays in lighted bedroom 1 for 2 hours, walks outside for 1 hour and then leaves to watch TV in the living room with light turned on for 3 hours, after which he turns off the TV and sleeps in bedroom 1 for 3 hours. When waking up, the light is automatically turned on in bedroom 1 for 1 hour. In this situation, the energy is saved from the 4 hours being out and sleeping. Meanwhile, resident Y is outside for 1 hour, then enters the living room for 1 hour with the light on. Afterwards, he goes to bedroom 2 and stays 2 hours with the light on, before he sleeps 2 hours, followed by waking up and watching TV for 4 hours with the light on. The energy saved here is 1 hour when he was outside, along with 2 hours of sleeping. The two concurrent situations are illustrated in Fig. 4.

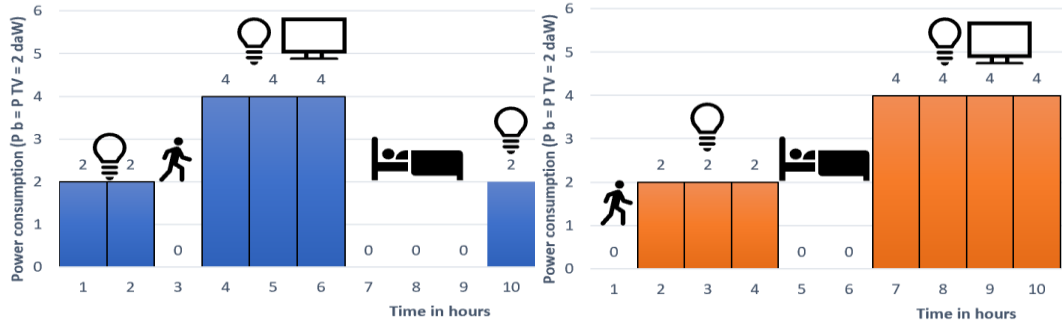

Fig. 4. Energy consumption for resident X in blue and Y in orange

Power consumption expressed in dekawatts for bulbs ($P_B$) and TVs ($P_{TV}$) is computed for both residents during the same 10 hours interval. The outcomes are illustrated in Fig. 5. The total consumption is equal to 40 daW/h (18 daW/h for resident X and 22 daW/h for resident Y).

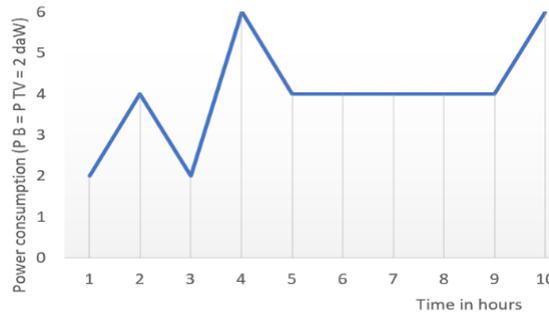

Fig. 5. Total power consumption based on the simulated 10 hours interval

Another scenario, illustrated in Fig. 6, is tested for the situation when besides the two residents of the house, there is another temporary visitor. For this situation, once the visitor enters the house, they all stay in the daylighted living room for 1 hour, then watch TV together for 3 hours.

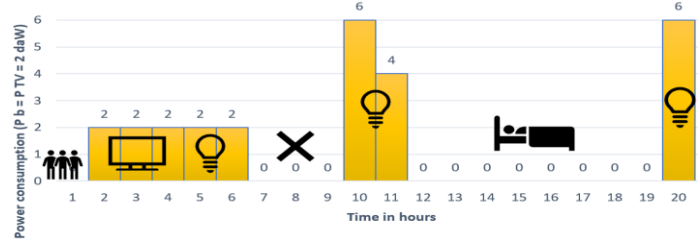

Fig. 6. Energy consumption for all the three persons

After watching TV, the agent turns on the light, as the room gets darken for 2 hours, before they all leave the smart home for 3 hours. Once they reach home, the lights of living room and of the two bedrooms are turned on for 1 hour.

Afterwards, for 1 hour, the visitor watches TV with resident X in bedroom 1 without light, while resident Y is in bedroom 2 with light on, also for 1 hour. After 8 hours during which residents X and Y, along with the visitor sleep, the light is turned on for 1 hour in the two bedrooms and the living room. The consumed energy for the 20 hours is equal to 26 daW/h.

### 4.2. Testing health factors

Sleeping patterns for residents are tested, based on similarity to the predefined values that are obtained from a practitioner. The system alerts if a resident sleeps less than 6 hours or above 9 hours per 24 hours interval.

The health factor related to toileting is determined, based on the predefined dataset from Table 1 that was validated by a group of doctors within a Romanian national project[4], and 8 tested residents' attributes from Table 2. Based on the dataset, the Euclidean distance is calculated for the KNN algorithm with K values of 3 and 5, along with KMC. The results of the tests are used to predict abnormalities, as described in Table 2.

*Table 1*

**Predefined dataset of age, weight, toileting and abnormality**

| Test Number | G_Age | G_Weight | G_Toileting | Abnormality |
|---|---|---|---|---|
| 1 | 44 | 6 | 12 | False |
| 2 | 43 | 9 | 14 | True |
| 3 | 45 | 5 | 9 | False |
| 4 | 48 | 6 | 2 | True |
| 5 | 52 | 6 | 9 | False |
| 6 | 53 | 3 | 8 | False |
| 7 | 53 | 5 | 16 | True |
| 8 | 59 | 6 | 3 | True |
| 9 | 61 | 6 | 10 | False |
| 10 | 63 | 4 | 8 | False |
| 11 | 68 | 9 | 3 | True |
| 12 | 66 | 8 | 9 | False |
| 13 | 66 | 6 | 14 | True |
| 14 | 73 | 6 | 16 | True |
| 15 | 77 | 5 | 7 | False |
| 16 | 75 | 4 | 10 | False |
| 17 | 79 | 5 | 15 | True |
| 18 | 83 | 8 | 18 | True |
| 19 | 85 | 5 | 9 | False |
| 20 | 83 | 8 | 17 | True |
| 21 | 81 | 4 | 3 | True |

---

[4] Smartcare Project, UEFISCDI 42PTE/2020, https://smartcare-project.eu/en/home-page/

*Table 2*

**Resident's values and abnormality results obtained by KNN for K=3, K=5 and KMC**

| Person (P_Age, P_Weight, P_Toileting) | Result for K=3 | Result for K=5 | Result for KMC |
|---|---|---|---|
| P1 (87,8,15) | True | True | True |
| P2 (66,5,9) | False | False | False |
| P3 (54,2,6) | False | False | False |
| P4 (77,4,14) | True | True | True |
| P5 (61,5,7) | False | False | False |
| P6 (70,7,7) | False | False | False |
| P7 (88,9,16) | True | True | True |
| P8 (61,8,2) | True | False | False |

Fig. 7 shows a 3D plotted graph of applying KNN algorithm on the 8 values of a resident. Dots in green are the cases where abnormality is false, while those in red are the true values of it. In addition, the orange dots are the tested cases.

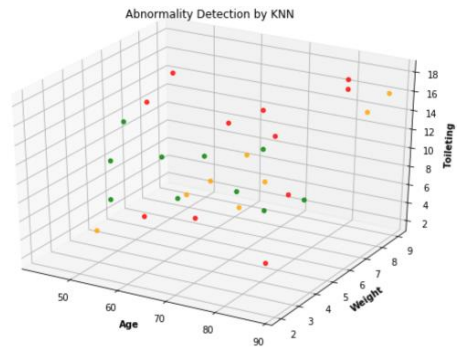

Fig. 7. Abnormality detection based on KNN

In Fig. 8 is the outcome obtained after applying the KMC algorithm. The centroids are fixed arbitrarily and after 10 iterations the clusters remain unchanged. The dark colored points belong to the cluster of normal values, while the light colored ones are for the abnormal values, based on the three considered criteria from Table 1. The results of applying KMC and KNN with K=3 vary for person P8. However, the results are identical both for KMC and KNN, when K is set to 5 in the KNN algorithm.

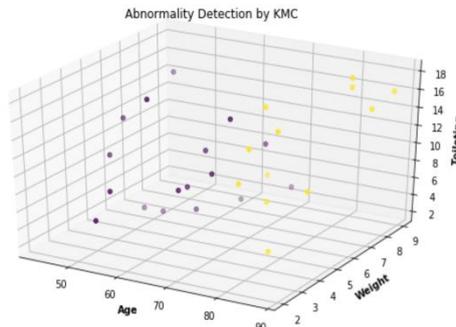

Fig. 8. Abnormality detection based on KMC

## 5. Conclusions

The purpose of this paper is to simulate a context-aware system of a smart home that manages power consumption. Managing power consumption is implemented to influence lights and TVs in a smart home. The environment is configured to save energy in cases of bright locations, sleeping in bedrooms, and leaving a location. Another purpose of this paper is to monitor some health issues. Factors of sleeping patterns and toileting frequency are considered. The used smart agents can alert once the sleeping pattern of a resident is not matching the predefined ones. Supervised and unsupervised machine learning techniques are introduced. For the toileting factor, the number of times using a toilet is counted and a machine learning technique is implemented to classify if the occupant requires health assistance.

The research will be improved in the future work on both energy and health factors. Regarding the first factor, more intelligent devices will be introduced to save energy, such as smart bulbs that can operate on certain brightness levels, and other devices, like smart shades that allow an amount of outside brightness to penetrate through windows and consequently reduce their operated time. Secondly, the health factor will be improved, based on human activity and pattern recognition. For this purpose, a mathematical statistical model, the Hidden Markov Model (HMM), will be applied to give probabilistic features for an activity to occur, which can help in predicting certain hidden cases. Also, human activity recognition (HAR) system, based on video camera surveillance, will be used to filter human actions and predict abnormalities. Finally, a data-mining framework will be implemented, based on the HMM and HAR models to better understand the residents' behavior.

## R E F E R E N C E S

[1] *C. H. Lim, P. Anthony and L. C. Fan*, "Applying Multi-Agent System in a Context Aware Smart Home", in Learning, **vol. 24**, 2009, pp.53-64.
[2] *S. Chang, and K. Nam*, "Smart Home Adoption: The Impact of User Characteristics and Differences in Perception of Benefits", in Buildings, **vol. 11**, no. 9, 2021, pp. 393.
[3] *M. Ragazzi, S. Fedrizzi, E. C. Rada, G. Ionescu, R. Ciudin and L. I. Cioca*, "Experiencing Urban Mining in an Italian Municipality towards a Circular Economy Vision", in Energy Procedia, **vol. 119**, 2017, pp. 192-200.
[4] *D. Nugent and B. K. Sovacool*, "Assessing the Lifecycle Greenhouse Gas Emissions from Solar PV and Wind Energy: A Critical Meta-survey", in Energy Policy, **vol. 65**, 2014, pp. 229-244.
[5] *A. Alberdi, A. Weakley, M. Schmitter-Edgecombe, D. Cook, A. Aztiria, A. Basarab and M. Barrenechea*, "Smart Home-Based Prediction of Multi-Domain Symptoms related to Alzheimer's Disease", in Energy Policy, **vol. 65**, 2014, pp. 229-244.
[6] *J. J. Borking, B. M. A. Van Eck, P. Siepel and P. J. A. Verhaar*, "Intelligent Software Agents and Privacy", The Hague: Registratiekamer, 1999.
[7] *H. S. Nwana and D. T. Ndumu*, "A Brief Introduction to Software Agent Technology", in Agent Technology, Springer, Berlin, Heidelberg, 1998, pp. 29-47.

[8] *M. Wooldridge and N. R. Jennings*, "Intelligent Agents: Theory and Practice", in The Knowledge Engineering Review, **vol. 10**, no. 2, 1995, pp. 115-152.

[9] *N. R. Jennings, T. J. Norman and P. Faratin*, "ADEPT: An Agent-Based Approach to Business Process Management", in ACM Sigmod Record, **vol. 27**, no. 4, 1998, pp. 32-39.

[10] *P. Stone and M. Veloso*, "Multiagent Systems: A Survey from a Machine Learning Perspective", in Autonomous Robots, **vol. 8**, no. 3, 2000, pp. 345-383.

[11] *K. P. Sycara*, "Multiagent systems", in AI magazine, **vol. 19**, no. 2, 1998, pp. 79.

[12] *F. Zambonelli, N. R. Jennings and M. Wooldridge*, "Developing Multiagent Systems: The Gaia Methodology", in Transactions on Software Engineering and Methodology, **vol. 12**, 2003.

[13] *E. Oliveira, K. Fischer and O. Stepankova*, "Multi-Agent Systems: Which Research for which Applications", in Robotics and Autonomous Systems, **vol. 27**, no. 1-2, 1999, pp. 91-106.

[14] *J. Palanca, E. Val, A. Garcia-Fornes, H. Billhardt, J. M. Corchado and V. Julián*, "Designing a Goal-Oriented Smart-Home Environment", in Information Systems Frontiers, 2018.

[15] *K. I. Benta, A. Hoszu, L. Văcariu and O. Creț*, 2009, "Agent Based Smart House Platform with Affective Control", in Proceedings of the Euro American Conference on Telematics and Information Systems: New Opportunities to increase Digital Citizenship, 2009, pp. 1-7.

[16] *V. Gopinath, A. Srija and C. N. Sravanthi*, "Re-design of Smart Homes with Digital Twins", in Journal of Physics: Conference Series, **vol. 1228**, no. 1, 2019, pp. 012031.

[17] *M. H. Kabir, M. R. Hoque, H. Seo and S. H. Yang*, "Machine Learning based Adaptive Context-Aware System for Smart Home Environment", in International Journal of Smart Home, **vol. 9**, no. 11, 2015, pp.55-62.

[18] *M. Li, and Y. Wu*, "Intelligent control system of smart home for context awareness" in the International Journal of Distributed Sensor Networks, **vol. 18**, no. 3, 2022, pp. 1-16.

[19] *J. Wu, F. Orlandi, T. AlSkaif, D. O'Sullivan, and S. Dev*, "A Semantic Web Approach to uplift Decentralized Household Energy Data" in the Sustainable Energy, Grids and Networks, **vol. 32**, 2022, pp. 1-12.

[20] *C. Casey and G. Kelley*, "Influential Article Review - Maximizing Smart Home Energy Management with Geodesic Acceleration and LevMar" in Modern Sciences Journal, **vol. 11**, no. 1, 2022, pp. 1-13.

[21] *D. Rodriguez-Gracia, J. Piedra-Fernandez, L. Iribarne, J. Criado, R. Ayala, J. Alonso-Montesinos, and C.-U. M. de las Mercedes,* "Microservices and Machine Learning Algorithms for Adaptive Green Buildings" in Sustainability Journal, **vol. 11**, 2019, pp. 1-8.

[22] *O. Munoz, A. Ruelas, P. Rosales, A. Acuna, A. Suastegui, and F. Lara*, "Design and Development of an IoT Smart Meter with Load Control for Home Energy Management Systems" in Sensors Journal, **vol. 22**, 2022, pp. 1-24.

[23] *A. Adamenko, A. Fedorenko, B. Nussbaum, and E. Schikuta*, "N2SkyC: User Friendly and Efficient Neural Network Simulation Fostering Cloud Containers" in Neural Processing Letters, **vol. 53**, 2021, pp. 1753-1772.

[24] *O. Holmala*, "Designing a Protocol Agnostic Rule Engine for a Cross-Domain Solution", Master's Thesis, 2019.

[25] *N. Li, H. Kong, Y. Ma, G. Gong and W. Huai*, "Human Performance Modeling for Manufacturing Based on an Improved KNN Algorithm" in the International Journal of Advanced Manufacturing Technology, **vol. 84**, no. 1, 2016, pp. 473-483.

[26] *A. Jasinska-Piadlo, R. Bond, P. Biglarbeigi, R. Brisk, P. Campbell, F. Browne and D. McEneaneny*, "Data-Driven versus a Domain-Led Approach to K-Means Clustering on an Open Heart Failure Dataset" in the International Journal of Data Science and Analytics, 2022, pp. 1-18.

[27] *C. Violán, A. Roso-Llorach, Q. Foguet-Boreu, M. Guisado-Clavero, M. Pons-Vigués, E. Pujol-Ribera and J. M. Valderas,* "Multimorbidity Patterns with K-Means Nonhierachical Cluster Analysis" in the BMC Family Practice, **vol. 19**, 2018, pp. 1-11.